\documentclass[twocolumn,prb,showpacs]{revtex4}
\usepackage{amsfonts}
\usepackage{epstopdf}
\usepackage{graphicx,color,psfrag,latexsym}
\usepackage{amssymb, amsmath}
\usepackage[usenames,dvipsnames]{xcolor}
\usepackage{comment}

\begin{document}

\title{ Current-induced magnetization dynamics in two magnetic insulators
separated by a normal metal }
\author{ Hans Skarsv{\aa }g$^{1}$, Gerrit E. W. Bauer$^{2,3}$ and Arne
Brataas$^{1}$}
\affiliation{$^1$Department of Physics, Norwegian University of Science and Technology,
NO-7491 Trondheim, Norway\\
$^2$Institute for Materials Research, Tohoku University, Sendai 980-8577,
Japan\\
$^3$Kavli Institute of NanoScience, Delft University of Technology, 2628 CJ
Delft, The Netherlands }
\date{\today}

\begin{abstract}
We study the dynamics of spin valves consisting of two layers of magnetic
insulators separated by a normal metal in the macrospin model. A current
through the spacer generates a spin Hall current that can actuate the
magnetization via the spin-transfer torque. We derive expressions for the
effective Gilbert damping and the critical currents for the onset of
magnetization dynamics including the effects of spin pumping that can be
tested by ferromagnetic resonance experiments. The current generates an
amplitude asymmetry between the in-phase and out-of-phase modes. We briefly
discuss superlattices of metals and magnetic insulators.
\end{abstract}


\maketitle

\section{Introduction}

Electric currents induce spin-transfer torques in heterogeneous or textured
magnetic systems.\cite{Brataas:nmat12} In this context, magnetic insulators
such as yttrium iron garnet (YIG) combined with normal metal contacts
exhibiting spin-orbit interactions, such as Pt, have recently attracted
considerable interest, both experimentally \cite%
{Kajiwara:nat10,Sandweg:apl10,Sandweg:prl11,Vilela-Leao:apl11,Burrowes:apl12,Rezende:apl12,Nakayama:prl13}
and theoretically.\cite%
{Xiao:prb10,Slonczewski:prb10,Xingtao:epl11,Xiao:prl12,Kapelrud:prl13,Yan-Ting:prb13,Xiao:xxx13}
Since the discovery of non-local exchange coupling and giant
magnetoresistance in spin valves, i.e., a normal metal sandwiched between
two ferromagnetic metals, these systems have been known to display rich
physics. Some of these effects, such as the dynamic exchange interaction,%
\cite{Tserkovnyak:revmod05} should also arise when the magnetic layers are
insulators. The spin Hall magnetoresistance (SMR) is predicted to be
enhanced in such spin valves, \cite{Yan-Ting:prb13} although experimental
realizations have not yet been reported. Here, we consider multilayer
structures with ferromagnetic but electrically insulating (FI) layers and
normal metal (N) spacers. In-plane electric currents applied to N generate
perpendicular spin currents via the spin Hall effect (SHE). When these spin
currents are absorbed at the N$|$FI interfaces, the ensuing spin-transfer
torques can induce magnetization dynamics and switching. We consider ground
state configurations in which the magnetizations are parallel or
antiparallel to each other. For thin magnetic layers, even small torques can
effectively modify the (Gilbert) damping, which can be observed as changes
in the line width of the ferromagnetic resonance (FMR) spectra. We employ
the macrospin model for the magnetization vectors that is applicable for
sufficiently strong and homogeneous magnetic fields, while extensions are
possible.\cite{Xiao:prl12,Kapelrud:prl13,Xiao:xxx13} Our results include the
observation of effective (anti)damping resulting from in-plane charge
currents in FI$|$N$|$FI trilayers, magnetic stability analysis in the
current-magnetic field parameter space and a brief analysis of the dynamics
for currents above the critical value. We also consider current-induced
effects in superlattices. Our paper is organized as follows. In Section \ref{s:model}, we present our model for a FI$|$N$|$FI spin valve including the
SHE spin current generation and spin pumping, modeled as additional torques
in the Landau-Lifshitz-Gilbert equation. We proceed to formulate the
linearized magnetization dynamics and the spin accumulation in N in Section~\ref{s:torques}. In Section~\ref{s:eigenmodes}, we calculate the eigenmodes
and the current-controlled effective Gilbert damping and determine the
critical currents at which the magnetic precession becomes unstable. We
discuss the current-induced dynamics of $\ \cdots |\mathrm{FI}|\mathrm{N}|%
\mathrm{FI}|\mathrm{N}|\cdots $ superlattices in Section~\ref{s:superlattice}%
. Finally, we summarize our conclusions and provide an outlook in Section %
\ref{s:conclusions}.

\section{Model}

\label{s:model} FI1$|$N$|$FI2 denotes the heterostructure composed of a
normal metal (N) layer sandwiched between two layers of ferromagnetic
insulators (FIs) (see Fig.~\ref{fig:parallel}). We denote the thicknesses of
FI1, N and FI2 by $d_{1}$, $d_{\text{\textrm{N}}}$ and $d_{2}$,
respectively. We adopt a macrospin model of spatially constant magnetization 
$\mathbf{M}_{i}$ in each layer. The magnetization dynamics of the two layers
are described by the coupled Landau-Lifshitz-Gilbert-Slonczewski (LLGS)
equations: 
\begin{eqnarray}
\mathbf{\dot{M}}_{i} &=&-\gamma \mathbf{M}_{i}\times \left( \mathbf{H}_{%
\mathrm{eff},i}+\frac{J}{d_{i}M_{S,i}}\mathbf{M}_{j}\right) +\alpha _{i}%
\mathbf{M}_{i}\times \mathbf{\dot{M}}_{i}  \notag  \label{eq:LLG} \\
&&+\boldsymbol{\tau }_{i}^{\mathrm{DSP}}+\boldsymbol{\tau }_{i}^{\mathrm{ISP}%
}+\boldsymbol{\tau }_{i}^{\mathrm{SH}},  \label{LLG}
\end{eqnarray}%
where $\mathbf{M}_{i}$ is the unit vector in the direction of the
magnetization in the left/right layer with indices $i=1,2$; $M_{S,i}$ is the
saturation magnetization; $\gamma $ is the gyromagnetic ratio; $\alpha _{i}$
is the Gilbert damping constant; $J$ is the interlayer dipolar and exchange
energy areal density, with $j=1(2)$ when $i=2(1)$; and $\mathbf{H}_{\mathrm{%
eff},i}$ is an effective magnetic field: 
\begin{equation}
\mathbf{H}_{\mathrm{eff},i}=\mathbf{H}_{\text{\textrm{ext}}}+\mathbf{H}_{%
\text{\textrm{an}},i}\left( \mathbf{M}_{i}\right)
\end{equation}%
consisting of the external magnetic field $\mathbf{H}_{\mathrm{ext}}$ as
well as the anisotropy fields $\mathbf{H}_{\mathrm{an},i}$ for the
left/right layer. We distinguish direct (DSP) and indirect spin pumping
(ISP). DSP generates the spin angular momentum current $\mathbf{j}_{1(2)}^{%
\mathrm{DSP}}$ through the interfaces of FI1(2). A positive spin current
corresponds to a spin flow toward the FI from which it originates. The DSP
spin current is expressed as 
\begin{equation}
\mathbf{j}_{i}^{\mathrm{DSP}}=\frac{\hbar }{e}g_{\perp ,i}\mathbf{M}%
_{i}\times \mathbf{\dot{M}}_{i},  \label{eq:spin-pumping current}
\end{equation}%
where $g_{\perp ,i}$ is the real part of the spin-mixing conductance of the N%
$|$FI1(2) interface per unit area for $i=1(2)$, respectively, and $-e$ is
the electron charge. This angular momentum loss causes a damping torque
(here and below in CGS units): 
\begin{equation}
\boldsymbol{\tau }_{i}^{\mathrm{DSP}}=\frac{\gamma \hbar ^{2}g_{\perp ,i}}{%
2e^{2}M_{S,i}d_{i}}\mathbf{M}_{i}\times \mathbf{\dot{M}}_{i}.
\label{eq:spin-pumping torque}
\end{equation}%
In ballistic systems, the spin current emitted by the neighboring layer is
directly absorbed and generates an indirect spin torque on the opposing
layer:\cite{Tserkovnyak:revmod05} 
\begin{equation}
\boldsymbol{\tau }_{i,\mathrm{ball}}^{\mathrm{ISP}}=-\frac{\gamma \hbar
^{2}g_{\perp ,i}}{2e^{2}M_{S,i}d_{i}}\mathbf{M}_{i}\times \mathbf{\dot{M}}%
_{i}.
\end{equation}%
In the presence of an interface or bulk disorder, the transport is diffuse,
and the ISP is 
\begin{equation}
\boldsymbol{\tau }_{i}^{\mathrm{ISP}}=-\frac{\gamma \hbar }{%
2e^{2}M_{S,i}d_{i}}g_{\perp ,i}\mathbf{M}_{i}\times (\mathbf{M}_{i}\times 
\boldsymbol{\mu }^{\mathrm{SP}}(z_{i})),  \label{eq:spin-transfer torque}
\end{equation}%
where $\boldsymbol{\mu }^{\mathrm{SP}}(z_{i})$ is the spin pumping
contribution to the spin accumulation (difference in chemical potentials) at
the interface in units of energy, with $z_{i}\equiv \mp d_{\mathrm{N}}/2$
for $i=1,2$. $\boldsymbol{\mu }^{\mathrm{SP}}$ is the solution of the spin
diffusion equation in N as discussed below.

Due to the SHE, an in-plane DC charge current produces a transverse spin
current that interacts with the FI$|$N interfaces. Focusing on the diffusive
regime, the areal density of charge current $\mathbf{j}_{\mathrm{c}}$ as
well as the spin $\mathbf{j}_{k}^{\mathrm{SH}}$ current$\ $in the $k$%
-direction, where $\mathbf{j}_{k}^{\mathrm{SH}}/\left\vert \mathbf{j}_{k}^{%
\mathrm{SH}}\right\vert $ is the spin polarization unit vector, can be
written in terms of a symmetric linear response matrix:\cite{Yan-Ting:prb13} 
\begin{eqnarray}
\left( 
\begin{matrix}
\mathbf{j}_{\mathrm{c}} \\ 
\mathbf{j}_{x}^{\mathrm{SH}} \\ 
\mathbf{j}_{y}^{\mathrm{SH}} \\ 
\mathbf{j}_{z}^{\mathrm{SH}}%
\end{matrix}%
\right) &=&\sigma \left( 
\begin{matrix}
1 & \Theta _{\mathrm{SH}}\hat{\mathbf{x}}\times & \Theta _{\mathrm{SH}}\hat{%
\mathbf{y}}\times & \Theta _{\mathrm{SH}}\hat{\mathbf{z}}\times \\ 
\Theta _{\mathrm{SH}}\hat{\mathbf{x}}\times & 1 & 0 & 0 \\ 
\Theta _{\mathrm{SH}}\hat{\mathbf{y}}\times & 0 & 1 & 0 \\ 
\Theta _{\mathrm{SH}}\hat{\mathbf{z}}\times & 0 & 0 & 1%
\end{matrix}%
\right)  \notag \\
&&\left( 
\begin{matrix}
-\nabla \mu _{c}/e \\ 
-\nabla \mu _{x}^{\mathrm{SH}}/(2e) \\ 
-\nabla \mu _{y}^{\mathrm{SH}}/(2e) \\ 
-\nabla \mu _{z}^{\mathrm{SH}}/(2e)%
\end{matrix}%
\right) ,  \label{sheishe}
\end{eqnarray}%
where $\Theta ^{\mathrm{SH}}$ is the spin Hall angle, $\sigma $ is the
electrical conductivity and $\mu _{c}$ is the charge chemical potential. $%
\boldsymbol{\mu }^{\mathrm{SH}}=(\mu _{x}^{\mathrm{SH}},\mu _{y}^{\mathrm{SH}%
},\mu _{z}^{\mathrm{SH}})$ is the spin accumulation induced by reflection of
the spin currents at the interfaces. The spin transfer torques $\boldsymbol{%
\tau }_{i}^{\mathrm{SH}}$ at the FI interfaces ($i=1,2$) are then expressed
as 
\begin{equation}
\boldsymbol{\tau }_{i}^{\mathrm{SH}}=-\frac{\gamma \hbar }{2e^{2}M_{S,i}d_{i}%
}g_{\perp ,i}\mathbf{M}_{i}\times \left( \mathbf{M}_{i}\times \boldsymbol{%
\mu }^{\mathrm{SH}}(z_{i})\right) .  \label{eq:STT}
\end{equation}%
The polarization of $\boldsymbol{\mu }^{\mathrm{SH}}$ and thereby $%
\boldsymbol{\tau }_{i}^{\mathrm{SH}}$ can be controlled by the charge
current direction. In the following sections, we assume that the shape
anisotropy and exchange coupling favor parallel or antiparallel equilibrium
orientations of $\mathbf{M}_{1}$ and $\mathbf{M}_{2}$. For small current
levels, the torques normal to the magnetization induce tilts from their
equilibrium directions and, at sufficiently large currents, trigger
complicated dynamics, while torques directed along the equilibrium
magnetization modify the effective damping and induce magnetization
reversal. Here, we focus on the latter configuration, in which the spin
accumulation in N is collinear to the equilibrium magnetizations.

In the following equations, we take the thickness, saturation magnetization,
Gilbert damping and spin-mixing conductance to be equal in the two layers
FI1 and FI2, with an out-of-plane hard axis and an in-plane internal field: 
\begin{subequations}
\begin{eqnarray}
\mathbf{H}_{\mathrm{eff},1} &=&\frac{\omega _{H}}{\gamma }\hat{\mathbf{x}}-%
\frac{\omega _{M}}{\gamma }(\mathbf{M}_{1})_{z}\hat{\mathbf{z}}, \\
\mathbf{H}_{\mathrm{eff},2} &=&s\frac{\omega _{H}}{\gamma }\hat{\mathbf{x}}-%
\frac{\omega _{M}}{\gamma }(\mathbf{M}_{2})_{z}\hat{\mathbf{z}},
\end{eqnarray}%
with $\omega _{H}=\gamma (H_{\text{\textrm{ext}}}+(H_{\mathrm{an},i})_{x})$
and $\omega _{M}=4\pi \gamma M_{S}$. Pure dipolar interlayer coupling with $%
J<0$ favors an antiparallel ground state configuration, while the exchange
coupling oscillates as a function of $d_{N}$. 
\begin{figure}[h]
\includegraphics[width=8cm]{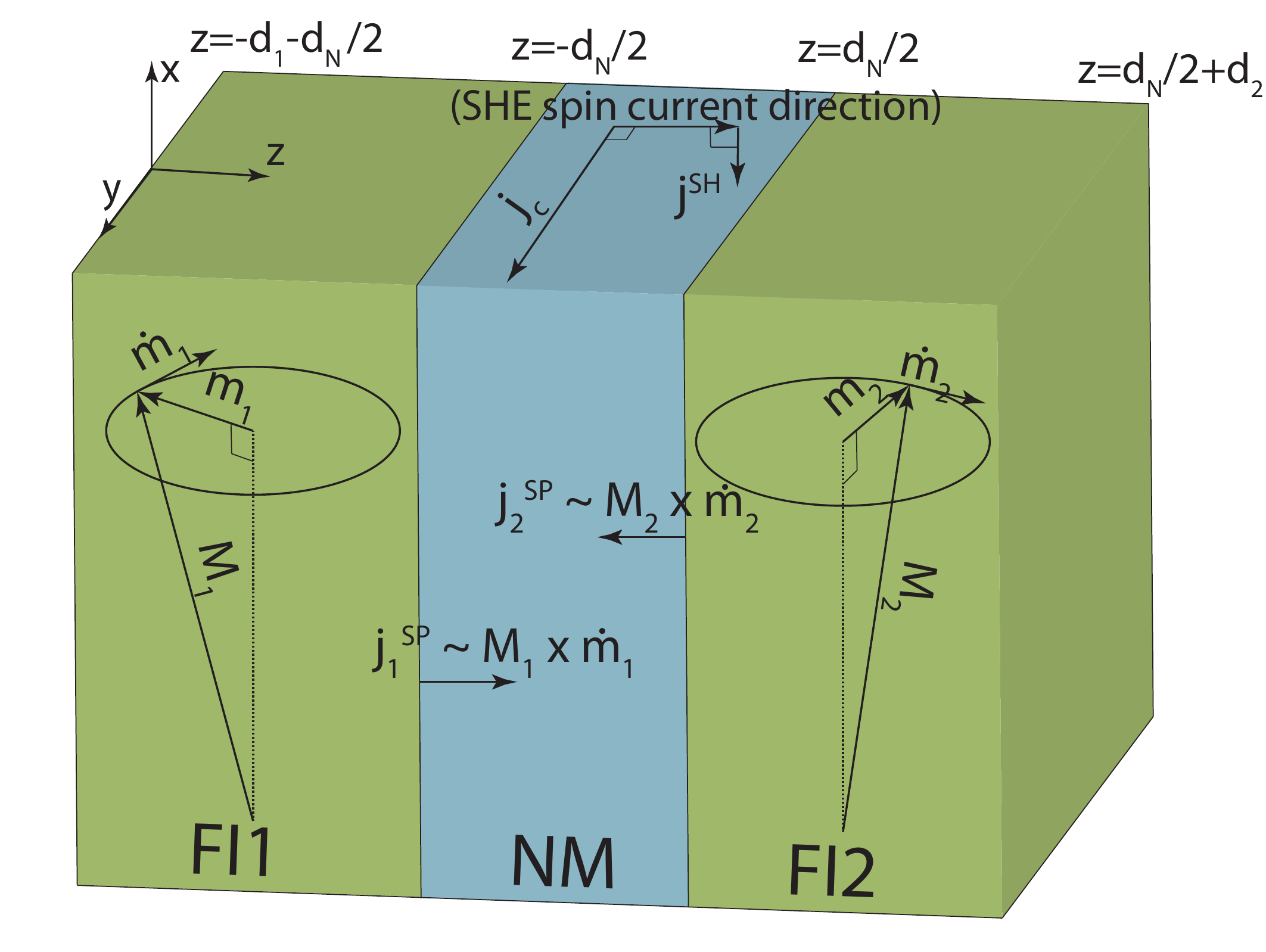}
\caption{(Color online) Spin valve of ferromagnetic insulators (FIs)
sandwiching a normal metal (N). The equilibrium magnetizations $\mathbf{M}%
_{1}$ and $\mathbf{M}_{2}$ are collinear, i.e., parallel or antiparallel. A
spin-Hall-induced spin current flows in the $z$-direction and is polarized
along $x$.\label{fig:parallel}}
\end{figure}

\section{Spin-transfer torques}\label{s:torques}

The spin-pumping and spin-transfer torques $\boldsymbol{\tau }_{i}^{\mathrm{%
DSP}}$ and $\boldsymbol{\tau }_{i}^{\mathrm{ISP}}$ (Eqs.~%
\eqref{eq:spin-pumping torque} and \eqref{eq:spin-transfer torque}) cause
dynamic coupling between the two magnetizations. To leading order, these
torques can be treated separately. We now derive expressions for disordered
systems that support spin accumulations $\boldsymbol{\mu }^{\mathrm{X}%
}\left( z\right) $ $(\mathrm{X}=\mathrm{SH},\mathrm{SP})$ governed by the
spin-diffusion equation: 
\end{subequations}
\begin{equation}
\boldsymbol{\dot{\mu}}^{\mathrm{X}}=D\partial _{z}^{2}\boldsymbol{\mu }^{%
\mathrm{X}}-\frac{\boldsymbol{\mu }^{\mathrm{X}}}{\tau _{\mathrm{sf}}}.
\label{eq:spin-pumping diffusion}
\end{equation}%
Here, $D$ is the diffusion constant, and $\tau _{\mathrm{sf}}$ is the
spin-flip relaxation time. The diffuse spin current in the $z$-direction
related to this spin accumulation follows Eq.~(\ref{sheishe}): 
\begin{equation}
\mathbf{j}^{\mathrm{X}}=-\frac{\sigma }{2e}\frac{\partial \boldsymbol{\mu }^{%
\mathrm{X}}}{\partial z},
\end{equation}%
where $\sigma $ is the conductivity of N.

\subsection{Spin-pumping-induced torques}

The total spin current into an FI is the sum of the spin-transfer and
spin-pumping currents. Disregarding interface spin-flip scattering, the
boundary conditions for the left/right layer are 
\begin{equation}
-\frac{1}{e}g_{\perp }\mathbf{M}_{i}\times \left( \mathbf{M}_{i}\times 
\boldsymbol{\mu }^{\mathrm{SP}}(z_{i})\right) +\mathbf{j}_{i}^{\mathrm{DSP}%
}=\mp \mathbf{j}^{\mathrm{SP}}\left( z_{i}\right) .
\label{eq:spin-pumping current conservation}
\end{equation}%
The -(+) sign on the right-hand side is due to the opposite flow direction
of the spin currents at the left (right) interface. We expand the
magnetization direction around the equilibrium configuration as 
\begin{subequations}
\begin{eqnarray}
\mathbf{M}_{1} &=&\mathbf{\hat{x}}+\mathbf{m}_{1}, \\
\mathbf{M}_{2} &=&s\mathbf{\hat{x}}+\mathbf{m}_{2},
\end{eqnarray}%
as long as $\left\vert \mathbf{m}_{i}\right\vert \ll \left\vert \mathbf{M}%
_{i}\right\vert $ or $\mathbf{m}_{i}\cdot \mathbf{M}_{i}=\mathcal{O}\left(
\left\vert \mathbf{m}_{i}\right\vert ^{2}\right) $. The parameter $s=1$ when
the equilibrium configuration is parallel; $s=-1$ when it is antiparallel.
The FMR frequency is usually much smaller than the diffuse electron
traversal rate $D/d_{\mathrm{N}}^{2}$ and spin-flip relaxation $1/\tau _{%
\mathrm{sf}}$ rate; thus, retardation of the spin flow may be disregarded.
In the steady state, the left-hand side of Eq.~%
\eqref{eq:spin-pumping
diffusion} vanishes. We solve Eq.~\eqref{eq:spin-pumping diffusion} for the
adiabatic magnetization dynamics with boundary conditions Eq.~%
\eqref{eq:spin-pumping current conservation} to obtain the spin
accumulation: 
\end{subequations}
\begin{eqnarray}
\boldsymbol{\mu }^{\mathrm{SP}} &=&-\frac{\hbar }{2}\mathbf{\hat{x}}\times %
\left[ (\mathbf{\dot{m}}_{1}+s\mathbf{\dot{m}}_{2})\Gamma _{1}\left(
z\right) \right.  \notag \\
&&\left. -(\mathbf{\dot{m}}_{1}-s\mathbf{\dot{m}}_{2})\Gamma _{2}\left(
z\right) \right] ,
\end{eqnarray}%
where $l_{\mathrm{sf}}=\sqrt{D\tau _{\mathrm{sf}}}$ is the spin-diffusion
length and 
\begin{subequations}
\begin{eqnarray}
\Gamma _{1}\left( z\right) &\equiv &\frac{\cosh \left( z/l_{\mathrm{sf}%
}\right) }{\cosh \left( z/l_{\mathrm{sf}}\right) +\sigma \sinh \left( z/l_{%
\mathrm{sf}}\right) /2g_{\perp }l_{\mathrm{sf}}}, \\
\Gamma _{2}\left( z\right) &\equiv &\frac{\sinh \left( z/l_{\mathrm{sf}%
}\right) }{\sinh \left( z/l_{\mathrm{sf}}\right) +\sigma \cosh \left( z/l_{%
\mathrm{sf}}\right) /2g_{\perp }l_{\mathrm{sf}}}.
\end{eqnarray}%
The torques are 
\end{subequations}
\begin{equation}
\boldsymbol{\tau }_{i}^{\mathrm{ISP}}=\frac{\gamma \hbar }{2e^{2}M_{S}d}%
g_{\perp }\boldsymbol{\mu }^{\mathrm{SP}}(z_{i}).
\label{eq:spin-transfer torque2}
\end{equation}%
Because the spin accumulation is generated by the dynamics of both
ferromagnets, we obtain spin-pumping-induced dynamic coupling that is
quenched when $d_{\mathrm{N}}\gg l_{\mathrm{sf}}$. In the limit of vanishing
spin-flip scattering, the spin accumulation is spatially constant and is
expressed as 
\begin{equation}
\boldsymbol{\mu }^{\mathrm{SP}}\overset{d_{\text{\textrm{N}}}\ll l_{\mathrm{%
sf}}}{\rightarrow }-\frac{\hbar }{2}\mathbf{\hat{x}}\times (\mathbf{\dot{m}}%
_{1}+s\mathbf{\dot{m}}_{2}).
\end{equation}%
The corresponding diffusive torque is then a simple average of the
contributions from the two spin-pumping currents, in contrast to the
ballistic torque that depends only on the magnetization on the opposite side.

\subsection{Current-induced torques}

A charge current in the $y$-direction causes a spin Hall current in the $z$%
-direction that is polarized along the $x$-direction (see Fig.~\ref{fig:parallel}). At the interfaces, the current induces a spin-accumulation $%
\boldsymbol{\mu }^{\mathrm{SH}}$ that satisfies the diffusion Eq.~\eqref{eq:spin-pumping diffusion} and drives a spin current (dropping the
index $z$ from now on): 
\begin{equation}
\mathbf{j}^{\mathrm{SH}}=-\frac{\sigma }{2e}\frac{\partial \boldsymbol{\mu }%
^{\mathrm{SH}}}{\partial z}-j_{0}^{\mathrm{SH}}\hat{\mathbf{x}},
\end{equation}%
where $j_{0}^{\mathrm{SH}}=\Theta _{\mathrm{SH}}j_{c}.\ $Angular momentum
conservation at the left/right boundaries leads to 
\begin{equation}
-\frac{1}{e}g_{\perp }\mathbf{M}_{i}\times \left( \mathbf{M}_{i}\times 
\boldsymbol{\mu }^{\mathrm{SH}}(z_{i})\right) =\mp \mathbf{j}^{\mathrm{SH}%
}(z_{i}).  \label{current conservation}
\end{equation}%
When $\mathbf{M}_{i}\parallel \boldsymbol{\mu }^{\mathrm{SH}}$, the spin
Hall current is completely reflected and the spin current at the interface
vanishes, while the absorption and torque are maximal when $\mathbf{M}%
_{i}\bot \boldsymbol{\mu }^{\mathrm{SH}}$. Spin currents and torques at the
interface scale favor $\mathbf{m}_{i}$ for small magnetization amplitudes.
Let us define a time-independent $\boldsymbol{\mu }_{0}^{\mathrm{SH}}$ for
collinear magnetizations and spin current polarization. For small dynamic
magnetizations, then 
\begin{equation}
\boldsymbol{\mu }^{\mathrm{SH}}=\boldsymbol{\mu }_{0}^{\mathrm{SH}}+\delta 
\boldsymbol{\mu }^{\mathrm{SH}},
\end{equation}%
where $\delta \boldsymbol{\mu }^{\mathrm{SH}}\sim \mathbf{m}_{i}$. We will show that the spin-Hall induced spin accumulation leads to a (anti)damping torque in the trilayer, while it gives a contribution to the real part of the frequency for superlattices (see Sec.~\ref{s:superlattice}).


Solving the diffusion Eq.~(\ref{eq:spin-pumping diffusion}) with boundary
conditions, Eq.~(\ref{current conservation}) yields 
\begin{equation}
\boldsymbol{\mu }_{0}^{\mathrm{SH}}=-\frac{2el_{\mathrm{sf}}}{\sigma }j_{0}^{%
\mathrm{SH}}\frac{\sinh (z/l_{\mathrm{sf}})}{\cosh (d_{\mathrm{N}}/2l_{%
\mathrm{sf}})}\mathbf{\hat{x}.}
\end{equation}%
The dynamic correction 
\begin{eqnarray}
\delta \boldsymbol{\mu }^{\mathrm{SH}} &=&-\frac{1}{2}\frac{2el_{\mathrm{sf}}%
}{\sigma }j_{0}^{\mathrm{SH}}\tanh (d_{\text{\textrm{N}}}/2l_{\mathrm{sf}}) 
\notag \\
&&\left[ (\mathbf{m}_{1}+s\mathbf{m_{2}})\Gamma _{2}(z)-(\mathbf{m}_{1}-s%
\mathbf{m_{2}})\Gamma _{1}(z)\right]  \label{eq:spin accumulation}
\end{eqnarray}%
leads to SHE torques [Eq.~(\ref{eq:STT})]: 
\begin{equation}
\boldsymbol{\tau }_{i}^{\mathrm{SH}}=-\frac{\gamma \hbar }{2e^{2}M_{S}d}%
g_{\perp }\left[ \mathbf{m}_{i}(\boldsymbol{\mu }_{0}^{\mathrm{SH}}\cdot 
\hat{\mathbf{x}})-\delta \boldsymbol{\mu }^{\mathrm{SH}}(z_{i})\right] .\ 
\label{SH coupling}
\end{equation}%
Eq.~\eqref{eq:LLG} then reduces to four coupled linear first-order partial
differential equations for $\mathbf{m}_{i}$.

\section{Eigenmodes and critical currents}

\label{s:eigenmodes} After linearizing Eq.~\eqref{LLG} and Fourier
transforming to the frequency domain $\dot{\mathbf{M}}_{i}\rightarrow
i\omega \hat{\mathbf{m}}_{i}$, Eq.~\eqref{eq:LLG} becomes 
\begin{equation}
\mathcal{M}\mathbf{v}=0,  \label{eq:matrix equation}
\end{equation}%
where $\mathbf{v}^{T}=(\hat{m}_{1,y},\hat{m}_{1,z},\hat{m}_{2,y},\hat{m}%
_{2,z})$ and $\mathcal{M}$ is a $4\times 4$ frequency-dependent matrix that
can be decomposed as 
\begin{equation}
\mathcal{M}=\mathcal{M}_{0}+J\mathcal{M}_{J}+(\alpha +\alpha ^{\prime })%
\mathcal{M}_{\mathrm{d}}+\alpha ^{\prime }\mathcal{M}_{\mathrm{SP}}+j_{0}^{%
\mathrm{SH}}\mathcal{M}_{\mathrm{SH}},
\end{equation}%
with 
\begin{subequations}
\label{eq:matrices}
\begin{eqnarray}
\mathcal{M}_{0} &=&\left( 
\begin{matrix}
-i\omega & -\tilde{\omega}_{H}-\omega _{M} & 0 & 0 \\ 
\tilde{\omega}_{H} & -i\omega & 0 & 0 \\ 
0 & 0 & -i\omega & -s\tilde{\omega}_{H}-s\omega _{M} \\ 
0 & 0 & s\tilde{\omega}_{H} & -i\omega%
\end{matrix}%
\right) , \\
\mathcal{M}_{\mathrm{d}} &=&\left( 
\begin{matrix}
0 & -i\omega & 0 & 0 \\ 
i\omega & 0 & 0 & 0 \\ 
0 & 0 & 0 & -is\omega \\ 
0 & 0 & is\omega & 0%
\end{matrix}%
\right) , \\
\mathcal{M}_{\mathrm{J}} &=&\left( 
\begin{matrix}
0 & 0 & 0 & \omega _{x} \\ 
0 & 0 & -\omega _{x} & 0 \\ 
0 & s\omega _{x} & 0 & 0 \\ 
-s\omega _{x} & 0 & 0 & 0%
\end{matrix}%
\right) , \\
\mathcal{M}_{\mathrm{ISP}} &=&\left( 
\begin{matrix}
0 & i\omega F^{\prime } & 0 & is\omega G^{\prime } \\ 
-i\omega F^{\prime } & 0 & -is\omega G^{\prime } & 0 \\ 
0 & i\omega G^{\prime } & 0 & is\omega F^{\prime } \\ 
-i\omega G^{\prime } & 0 & -is\omega F^{\prime } & 0%
\end{matrix}%
\right) , \\
\mathcal{M}_{\mathrm{SH}} &=&\left( 
\begin{matrix}
-F & 0 & -sG & 0 \\ 
0 & -F & 0 & -sG \\ 
G & 0 & sF & 0 \\ 
0 & G & 0 & sF%
\end{matrix}%
\right) .
\end{eqnarray}%
Here, $\mathcal{M}_{0}$ describes dissipationless precession in the
effective magnetic fields, and $\mathcal{M}_{\text{\textrm{d}}}$ arises from
Gilbert damping and the direct effect of spin pumping with a renormalized
damping coefficient $\tilde{\alpha}=\alpha +\alpha ^{\prime }$ and 
\end{subequations}
\begin{equation}
\alpha ^{\prime }=\frac{\gamma \hbar ^{2}}{2e^{2}M_{s}d}g_{\perp },
\label{eq:spdamping}
\end{equation}%
$\mathcal{M}_{\mathrm{J}}$ represents interlayer exchange coupling, $%
\mathcal{M}_{\mathrm{ISP}}$ represents spin-pumping-induced spin transfer,
and $\mathcal{M}_{\mathrm{SH}}$ represents the spin transfer caused by the
spin Hall current. The external and possible in-plane anisotropy fields are
modified by the interlayer coupling, $\omega _{H}\rightarrow \tilde{\omega}%
_{H}=\omega _{H}+\omega _{x}$, where $\omega _{x}=\gamma J/\left(
M_{s}d\right) $. The matrix elements $F^{\prime }$, $G^{\prime }$, $F$ and $%
G $ are generalized susceptibilities extracted from Eqs.~%
\eqref{eq:spin-transfer torque2} and \eqref{SH coupling}: 
\begin{subequations}
\begin{eqnarray}
F^{\prime } &=&\frac{1}{\alpha ^{\prime }}\frac{\partial (\boldsymbol{\tau }%
_{1}^{\mathrm{ST}})_{y}}{\partial \dot{m}_{1,z}}, \\
G^{\prime } &=&\frac{1}{\alpha ^{\prime }}\frac{\partial (\boldsymbol{\tau }%
_{1}^{\mathrm{ST}})_{y}}{\partial (s\dot{m}_{2,z})}, \\
F &=&-\frac{1}{j_{0}^{\mathrm{SH}}}\frac{\partial (\boldsymbol{\tau }_{1}^{%
\mathrm{ST}})_{y}}{\partial m_{1,y}}, \\
G &=&\frac{1}{j_{0}^{\mathrm{SH}}}\frac{\partial (\boldsymbol{\tau }_{1}^{%
\mathrm{ST}})_{y}}{\partial (sm_{2,y})}.
\end{eqnarray}%
The explicit expressions given in Appendix \ref{App: matrix elements} are
simplified for very thick and thin N spacers.

\textit{Thin N layer:} When $d_{\text{\textrm{N}}}\ll l_{\mathrm{sf}}\ $,
the interlayer coupling $G^{\prime }$ due to spin pumping approaches $%
F^{\prime},$, the intralayer coupling: 
\end{subequations}
\begin{equation}
G^{\prime }\rightarrow F^{\prime }\rightarrow \frac{1}{2},
\end{equation}
which implies that the incoming and outgoing spin currents are the same.
This outcome represents the limit of strong dynamic coupling in which the
additional Gilbert damping due to spin pumping vanishes when the
magnetization motion is synchronized.\cite{Heinrich:prl03} In this regime,
the SHE becomes ineffective because $F$ and $G $ scale as $d_{\text{\textrm{N%
}}}/l_{\mathrm{sf}}$. $F/G\rightarrow 2$ because $F$ contains a contribution
from both the static as well as the dynamic spin accumulation.

\textit{Thick N layer:} In the thick film limit, $d_{\text{\textrm{N}}}\gg
l_{\mathrm{sf}}$, the interlayer coupling vanishes as $G\rightarrow 0$ and $%
G^{\prime }\rightarrow 0$, while 
\begin{subequations}
\begin{eqnarray}
F^{\prime } &\rightarrow &\frac{1}{1+\frac{\sigma }{2g_{\perp }l_{\mathrm{sf}%
}}}, \\
F &\rightarrow &\frac{\gamma \hbar }{2eM_{S}d}\frac{1}{1+\frac{\sigma }{%
2g_{\perp }l_{\mathrm{sf}}}}.
\end{eqnarray}
Introducing the spin conductance $G_{\text{\textrm{sf}}}\equiv \mathcal{A}%
\sigma /2l_{\text{\textrm{sf}}}$ $G_{\perp }=\mathcal{A}g_{\perp }$ and $R_{%
\text{\textrm{tot}}}=(G_{\perp }+G_{\text{\textrm{sf}}})^{-1}$, the total
resistance of the interface and the spin active region of N, $F^{\prime
}\rightarrow R_{\text{\textrm{tot}}}G_{\perp}$, represents the backflow of
pumped spins. The same holds for the part of $F$ that originates from the
dynamic part of $\boldsymbol{\mu}^{\mathrm{SH}}$, while the static part
approaches a constant value when $d_{\mathrm{N}}$ becomes large (see
Appendix \ref{App: matrix elements}). In this limit, the system reduces to
two decoupled FI$|$N bilayers.

The eigenmodes of the coupled system are the solutions of $\det \left[ 
\mathcal{M}\left( \omega _{n}\right) \right] =0$ with complex
eigenfrequencies $\omega _{n}$. The SHE spin current induces spin
accumulations with opposite polarizations at the two interfaces. In the
parallel case, the torques acting on the two FIs are exerted in opposite
directions. The torques then stabilize one magnetization, but destabilize
the other. When the eigenfrequencies acquire a negative imaginary part,
their amplitude grows exponentially in time. We define the threshold current 
$j_{0,\mathrm{thr}}^{\mathrm{SH}}$ by the value at which $\text{Im}[\omega
_{n}\left( j_{0,\mathrm{thr}}^{\mathrm{SH}}\right)] =0.$ Because the total
damping has to be overcome at the threshold, $j_{0,\mathrm{thr}}^{\mathrm{SH}%
}\sim \tilde{\alpha}$. We treat the damping and exchange coupling
perturbatively, thereby assuming $\tilde{\alpha}\ll 1$ and $\omega _{x}\ll
\omega _{0},$ where $\omega _{0}=\sqrt{\tilde{\omega}_{H}(\tilde{\omega}%
_{H}+\omega _{M})}$ is the FMR frequency. The spin Hall angle is usually
much smaller than unity; thus, $j_{0}^{\mathrm{SH}}$ is treated as a
perturbation for currents up to the order of the threshold current, implying
that $\left\vert \text{Im}[\omega _{n}\left( j_{0}^{\mathrm{SH}}\right)]
\right\vert \ll \left\vert \text{Re}[\omega _{n}\left( j_{0}^{\mathrm{SH}%
}\right) ]\right\vert $.

The exchange coupling $\omega_{x}=\gamma J/\left( M_{s}d\right) $ for YIG$|$%
Pt$|$ YIG should be weaker than that of the well-studied metallic magnetic
monolayers, where it is known to become very small for $d\gtrsim 3~\mathrm{nm%
}$.\cite{Qiu:rapid92} In the following sections, we assume that $\omega
_{x}\ll \omega _{M}$ may be treated as a perturbation.

To treat the damping, spin pumping, spin-Hall-induced torques and static
exchange perturbatively, we introduce the smallness parameter $\epsilon $
and let $\alpha \rightarrow \epsilon \alpha $, $\alpha ^{\prime }\rightarrow
\epsilon \alpha ^{\prime }$, $j_{0}^{\mathrm{SH}}\rightarrow \epsilon j_{0}^{%
\mathrm{SH}}$, $\omega _{x}\rightarrow \epsilon \omega _{x}$. In the
following sections, a first-order perturbation is applied by linearizing in $%
\epsilon $ and subsequently setting $\epsilon =1$.

We transform $\mathcal{M}$ by the matrix $\mathcal{U}$ that diagonalizes $%
\mathcal{M}_{0}$ with eigenvalues $(\omega _{0},\omega _{0},-\omega
_{0},-\omega _{0})$. We then extract the part corresponding to the real
eigenfrequencies, which yields the following equation: 
\end{subequations}
\begin{equation}
\left\vert 
\begin{matrix}
(D)_{11} & (D)_{12} \\ 
(D)_{21} & (D)_{22}%
\end{matrix}%
\right\vert =0,
\end{equation}%
where $D=\mathcal{U}^{-1}\mathcal{M}\mathcal{U}$. We thus reduce the
fourth-order secular equation in $\omega $ to a second-order expression. To
the first order, we find for the parallel ($s=1$) case, 
\begin{equation}
\omega ^{\mathrm{P}}=\tilde{\omega}_{0}+i\frac{\alpha _{\mathrm{eff}}^{%
\mathrm{P}}}{2}(2\tilde{\omega}_{H}+\omega _{M}),  \label{eq:eigenvalues}
\end{equation}%
where we introduced a current-controlled effective Gilbert damping: 
\begin{eqnarray}
\alpha _{\mathrm{eff}}^{\mathrm{P}} &=&\alpha +\alpha ^{\prime }(1-F^{\prime
})  \notag \\
&&\pm \sqrt{\left( \alpha ^{\prime }G^{\prime }-i\frac{\omega _{x}}{\omega
_{0}}\right) ^{2}+\frac{4(F^{2}-G^{2})\left( j_{0}^{\mathrm{SH}}\right) ^{2}%
}{(2\tilde{\omega}_{H}+\omega _{M})^{2}}}.  \label{eq: parallel damping}
\end{eqnarray}%
The imaginary part of the square root in Eq.~\eqref{eq: parallel damping}
causes a first-order real frequency shift that we may disregard, i.e., $%
\text{Re}\left[ \omega ^{\mathrm{P}}\right] \approx \tilde{\omega}%
_{0}\approx \omega _{0}$. We thus find two modes with nearly the same
frequencies but different effective broadenings.

The critical current $j_{0,\mathrm{thr}}^{\mathrm{SH,P}}$ is now determined
by requiring that $\alpha _{\mathrm{eff}}^{\mathrm{P}}\ $ vanish, leading to 
\begin{eqnarray}
j_{0,\mathrm{thr}}^{\mathrm{SH,P}} &=&\pm \frac{\sqrt{\left( \alpha +\alpha
^{\prime }\left( 1-F^{\prime }\right) \right) ^{2}-\left( \alpha ^{\prime
}G^{\prime }\right) ^{2}}}{2\sqrt{F^{2}-G^{2}}}  \notag \\
&&\sqrt{1+\left( \frac{\omega _{x}/\omega _{0}}{\alpha +\alpha ^{\prime
}(1-F^{\prime })}\right) ^{2}}(2\tilde{\omega}_{H}+\omega _{M}),
\end{eqnarray}%
while the critical charge current is $j_{\mathrm{c},\mathrm{thr}}^{\mathrm{P}%
}=j_{0,\mathrm{thr}}^{\mathrm{S,P}}/\Theta _{\mathrm{SH}}$. Spin pumping and
spin flip dissipate energy, leading to a higher threshold current, which is
reflected by $1-F^{\prime }\geq G^{\prime }$. The reactive part of the
SHE-induced torque ($G$) suppresses the effect of the applied current and
thereby increases the critical current as well. The static exchange couples $%
\mathbf{M}_{1}$ and $\mathbf{M}_{2}$, hence increasing $j_{0,\mathrm{thr}}^{%
\mathrm{SH,P}}$. The critical spin current decreases monotonically with
increasing $d_{\mathrm{N}}/l_{\text{\textrm{sf}}}$, implying that the spin
valve (with parallel magnetization) has a larger threshold current than the
FI$|$N bilayer (with thick $d_{\mathrm{N}}$).

Analogous to the parallel case, we find two eigenmodes for the antiparallel
case ($s=-1$), with eigenfrequencies 
\begin{equation}
\omega ^{\mathrm{AP}}=\omega _{0}+\left( \pm \frac{-\omega _{x}}{2\omega _{0}%
}+i\frac{\alpha _{\mathrm{eff}}^{\mathrm{AP}}}{2}\right) (2\tilde{\omega}%
_{H}+\omega _{M})  \label{eq:eigenvalues antisymmetric}
\end{equation}%
and corresponding effective Gilbert damping parameters 
\begin{eqnarray}
\alpha _{\mathrm{eff}}^{\mathrm{AP}} &=&\alpha +\alpha ^{\prime
}(1-F^{\prime })  \notag \\
&&\pm \alpha ^{\prime }G^{\prime }\frac{\omega _{M}}{2\tilde{\omega}%
_{H}+\omega _{M}}+\frac{2}{2\tilde{\omega}_{H}+\omega _{M}}Fj_{0}^{\mathrm{SH%
}},
\end{eqnarray}%
which depend on the magnetic configuration because the dynamic exchange
coupling differs, while the resonance frequency is affected by the static
coupling. In the AP configurations, the spin Hall current acts with the same
sign on both layers due to the increase/decrease in damping on both sides
depending on the applied current direction. The corresponding threshold
current is expressed as 
\begin{equation}
j_{0,\mathrm{thr}}^{\mathrm{SH,AP}}=-\frac{\left( \alpha +\alpha ^{\prime
}(1-F^{\prime })\right) (2\tilde{\omega}_{H}+\omega _{M})-\alpha ^{\prime
}G^{\prime }\omega _{M}}{2F},
\end{equation}%
with $j_{c,\mathrm{thr}}^{\mathrm{SH,AP}}=j_{0,\mathrm{thr}}^{\mathrm{SH,AP}%
}/\Theta _{\mathrm{SH}}$. Again, the threshold for current-induced
excitation is increased by the spin pumping.

\begin{figure}[h]
\includegraphics[width=8cm,trim=2cm 2cm -1cm 8cm]{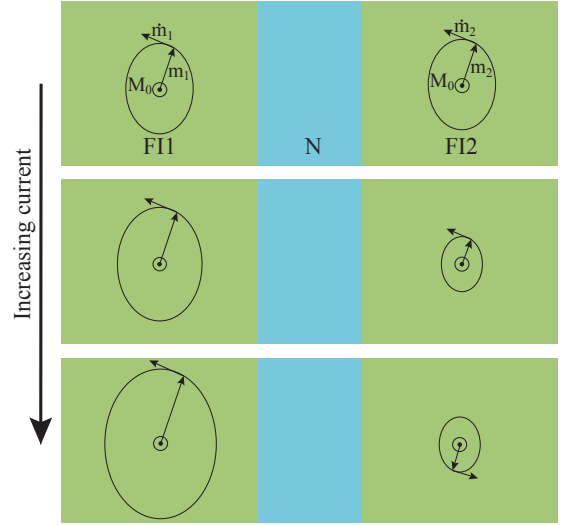}
\caption{(Color online) The acoustic mode for the parallel case for
different applied currents, ranging from zero to just below the critical current. For large currents the oscillations of
the two FIs become out of phase.}
\label{fig:acoustic}
\end{figure}
To zeroth order in the smallness parameter $\epsilon $, we find that the
eigenvectors for the parallel configuration take the form $\mathbf{v}^{%
\mathrm{P}}=(\mathbf{u},\beta \mathbf{u})^{T}$, where $\mathbf{u}$ is the
2-component vector 
\begin{equation}
\mathbf{u}=\left( 
\begin{matrix}
i\sqrt{1+\omega _{M}/\tilde{\omega}_{H}} \\ 
1%
\end{matrix}%
\right) .
\end{equation}%
The imbalance in the amplitudes of both layers is parameterized by 
\begin{equation}
\beta =\frac{2j_{0}^{\mathrm{SH}}F\mp \sqrt{4(F^{2}-G^{2})(j_{0}^{\mathrm{SH}%
})^{2}+\left( \alpha ^{\prime }G^{\prime }-i\frac{\omega _{x}}{\omega _{0}}%
\right) ^{2}(2\tilde{\omega}_{H}+\omega _{M})^{2}}}{-2j_{0}^{\mathrm{SH}%
}G+\left( \alpha ^{\prime }G^{\prime }-i\frac{\omega _{x}}{\omega _{0}}%
\right) (2\tilde{\omega}_{H}+\omega _{M})},
\end{equation}%
where $\mp $ corresponds to the $\pm $ in Eq.~\eqref{eq:eigenvalues}. For
the symmetric case, the applied current favors out-of-phase oscillations. It
can be demonstrated that in the limit of large currents and low spin-memory
loss, the corresponding amplitude difference is $\beta =-1$, with $j_{0}^{%
\mathrm{SH}}=0$, and an interlayer coupling dominated by either dynamic or
static exchange $\beta =\mp 1$, which correspond to an optical and an
acoustic mode, respectively. We use the labels \textquotedblleft
acoustic\textquotedblright\ and \textquotedblleft optic\textquotedblright
even though the phase difference is not precisely 0 or $\pi $ due to the
static exchange interaction. Note that $\beta (-j_{0}^{\mathrm{SH}})=1/\beta
(j_{0}^{\mathrm{SH}})$ is required by symmetry; inverting the current
direction is equivalent to interchanging FI1 and FI2. For $\omega _{x}=0$, $%
\beta (j_{0}^{\mathrm{SH}})$ is a pole or node depending on the current
direction for the acoustic mode in which the magnetization in one layer
vanishes. Above this current, $\beta $ change signs, and both modes have a
phase difference of $\pi $. The critical current lies above the current
corresponding to the node at which the acoustic mode becomes unstable. The
ballistic model also supports acoustic and optical modes, \cite%
{Tserkovnyak:revmod05} with the optical mode being more efficiently damped.

In the antiparallel case, acoustic and optical modes can are characterized
by amplitudes 
\begin{equation}
\mathbf{v}_{\mathrm{A}}^{\mathrm{AP}}=\left( 
\begin{matrix}
i\frac{\omega _{0}}{\tilde{\omega}_{H}} \\ 
1 \\ 
i\frac{\omega _{0}}{\tilde{\omega}_{H}} \\ 
-1%
\end{matrix}%
\right) ,\hspace{3mm}\mathbf{v}_{\mathrm{O}}^{\mathrm{AP}}=\left( 
\begin{matrix}
i\frac{\omega _{0}}{\tilde{\omega}_{H}} \\ 
1 \\ 
-i\frac{\omega _{0}}{\tilde{\omega}_{H}} \\ 
1%
\end{matrix}%
\right) ,  \label{eq:antiparModes}
\end{equation}%
where the optical (acoustic) mode corresponds to the {+(-)} sign in Eq.~%
\eqref{eq:eigenvalues antisymmetric}. The labels optical and acoustic are
kept because of the difference in effective damping; a 180$^{\circ }$
rotation about the $y$ axis of FI2 map these modes to the corresponding
modes for the parallel case. 
\begin{figure}[h]
\includegraphics[width=8cm,trim=1cm 15cm 1cm 1cm]{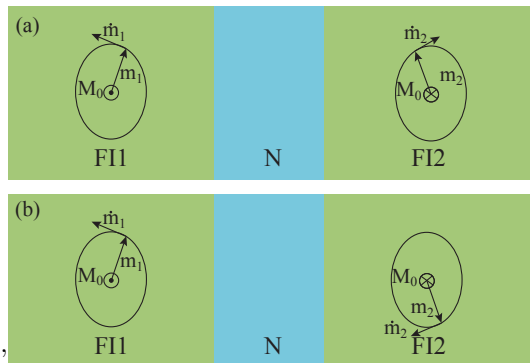}
\caption{(Color online) The eigenmodes of the antiparallel configuration.
(a)/(b) corresponds to the acoustic/optical mode of Eq.~ 
\eqref{eq:antiparModes}. For the acoustic/optical mode the
in-plane/out-of-plane component is equal in the two layers, and opposite for
the out-of-plane/in-plane component. }
\label{fig:antiparModes}
\end{figure}

\begin{table}[h!]
\caption{Physical parameters used in the numerical calculations}
\label{tab:constants}\centering
\par
\begin{tabular}{@{}lcc}
\hline
Constant & Value & Units \\ \hline
$g_\perp$ & $^a3.4 \cdot 10^{15}$ & $\mathrm{cm}^{-2}e^2/h$ \\ 
$\sigma$ & $^b5.4\cdot 10^{17}$ & $\mathrm{s}^{-1}$ \\ 
$4\pi M_S$ & $^c1750$ & $\mathrm{G}$ \\ 
$H_\text{int}$ & $0.2\cdot 4\pi M_S$ & $\mathrm{G}$ \\ 
$\alpha$ & $^c3 \cdot 10^{-4}$ &  \\ 
$l_\text{sf}$ & 10 & nm \\ 
$d_1, d_\text{N}, d_2$ & 10,~5,~10 & nm \\ \hline\hline
&  & 
\end{tabular}%
\par
\vspace{1mm}
\par
a) Ref.~[\onlinecite{Junfleisch:apl13}], b) Ref.~[%
\onlinecite{Giancoli:book84}], c) Ref.~[\onlinecite{Serga:jphysd}]
\end{table}
When the composition of the spin valve is slightly asymmetric, the dynamics
of the two layers can still be synchronized by the static and dynamic
coupling. However, at some critical detuning $\Delta \omega
=\omega_{2}-\omega _{1}$, this technique no longer works, as illustrated by
the eigenfrequencies for the asymmetric spin valve in Fig. \ref%
{fig:synchronization}. Here, we employ YIG$|$Pt$|$YIG parameters but tune
the FMR\ frequency of the right YIG layer. In practice, the tuning can be
achieved by varying the direction of the applied magnetic field.\cite%
{Heinrich:prl03} When the FMR frequencies of the two layers are sufficiently
close, the precessional motions in the two layers lock to each other. The
asymmetry introduced by higher currents is observed to suppress the
synchronization.

\begin{figure}[h]
\includegraphics[width=8cm]{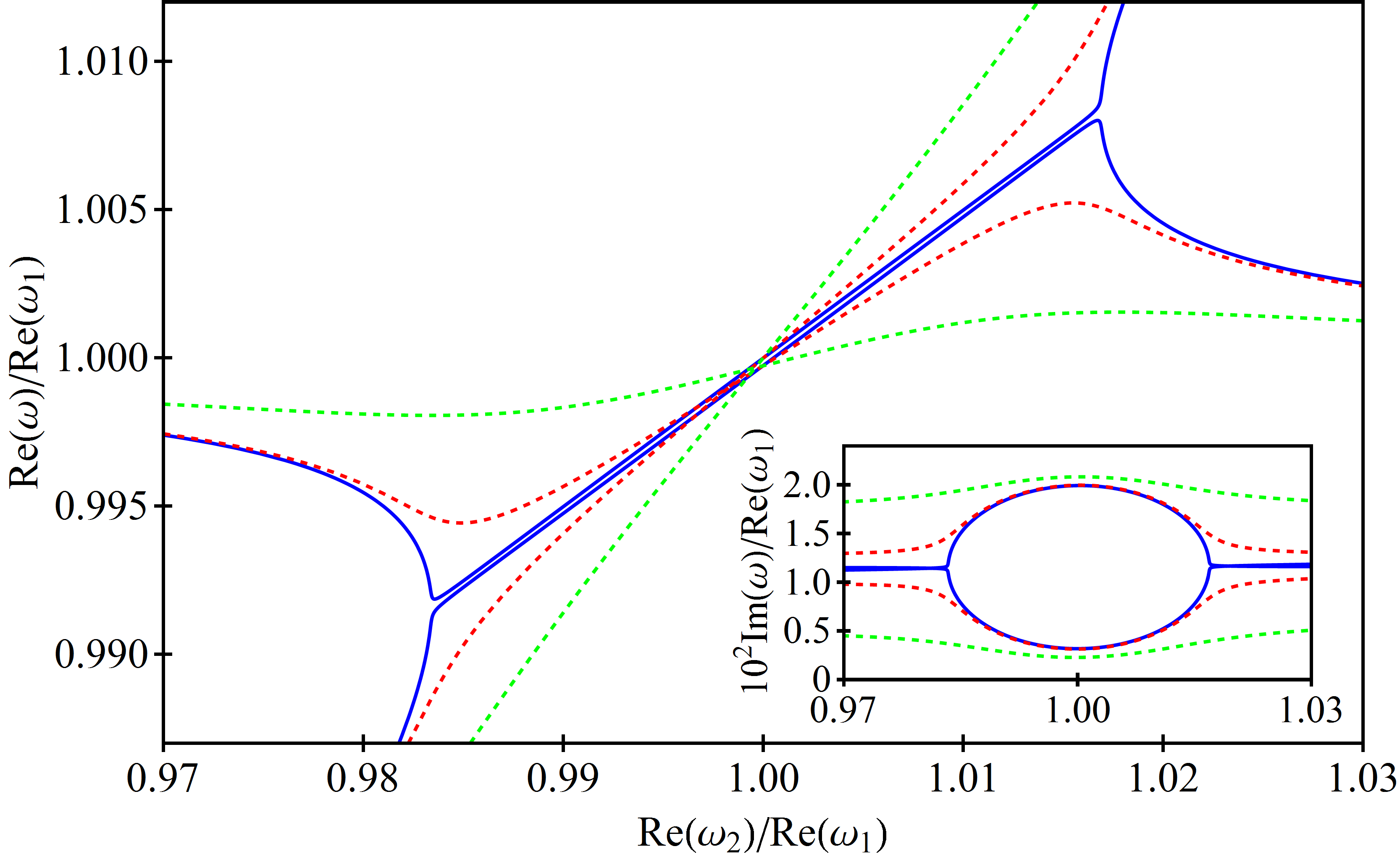}
\caption{(Color online) The lowest resonance frequencies of a parallel FI1$|$%
N$|$FI2 spin valve as a function of the detuning of the FMR frequencies of
the individual layers and for different currents $j_{0}^{\text{\textrm{SH}}%
}/j_{0,\text{\textrm{thr}}}^{\text{\textrm{SH}}}=0,10\%,50\%$ for the solid,
black dashed and grey dashed lines, respectively. At zero applied current
the two layers lock when detuning is small. The current suppresses
synchronization almost completely when reaching the threshold value.The
inset shows the corresponding broadenings.}
\label{fig:synchronization}
\end{figure}

\begin{figure}[h]
\label{fig:omegaparallel}\includegraphics[width=8cm]{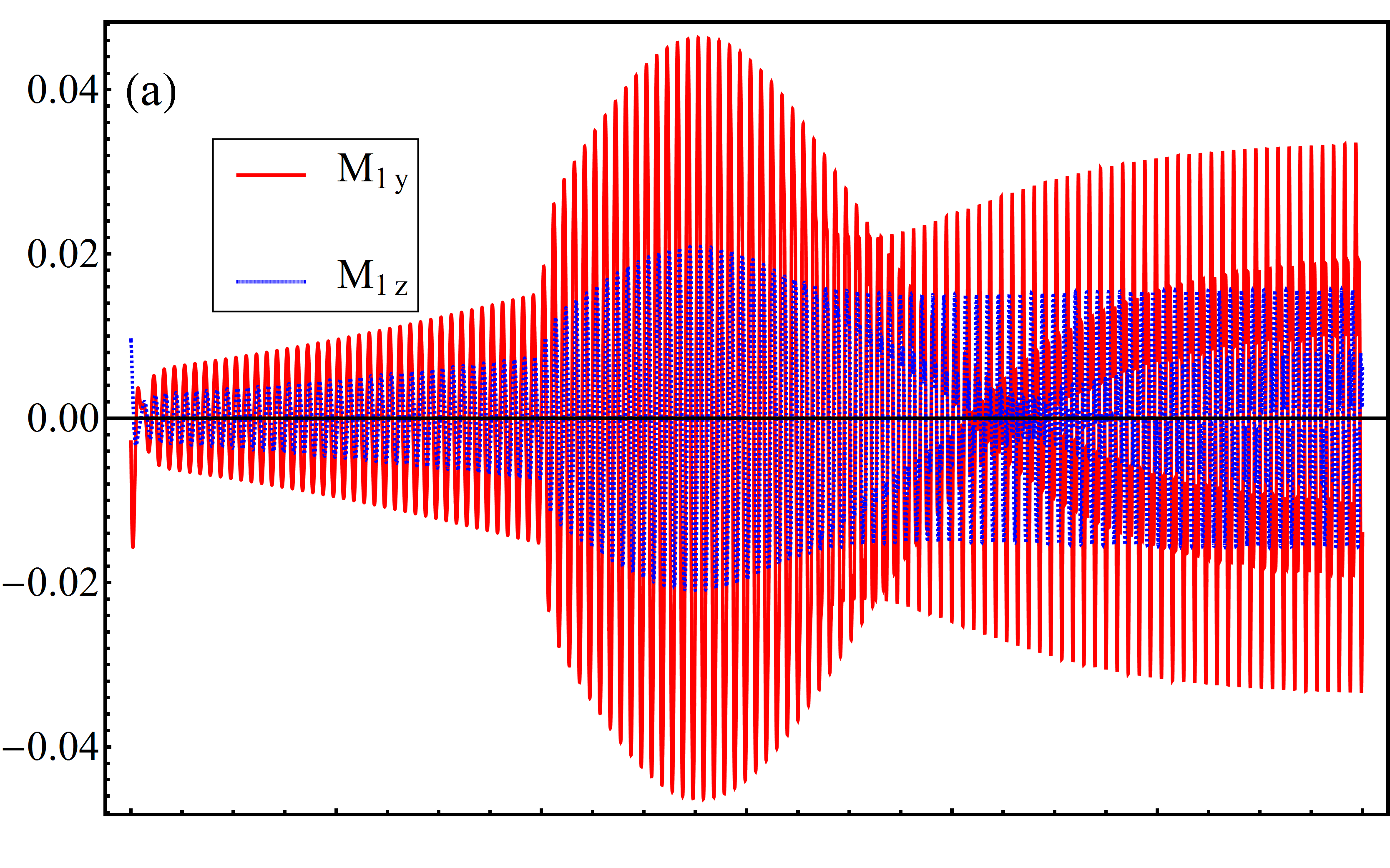}
\newline
\hspace{0.13cm}\includegraphics[width=7.87cm]{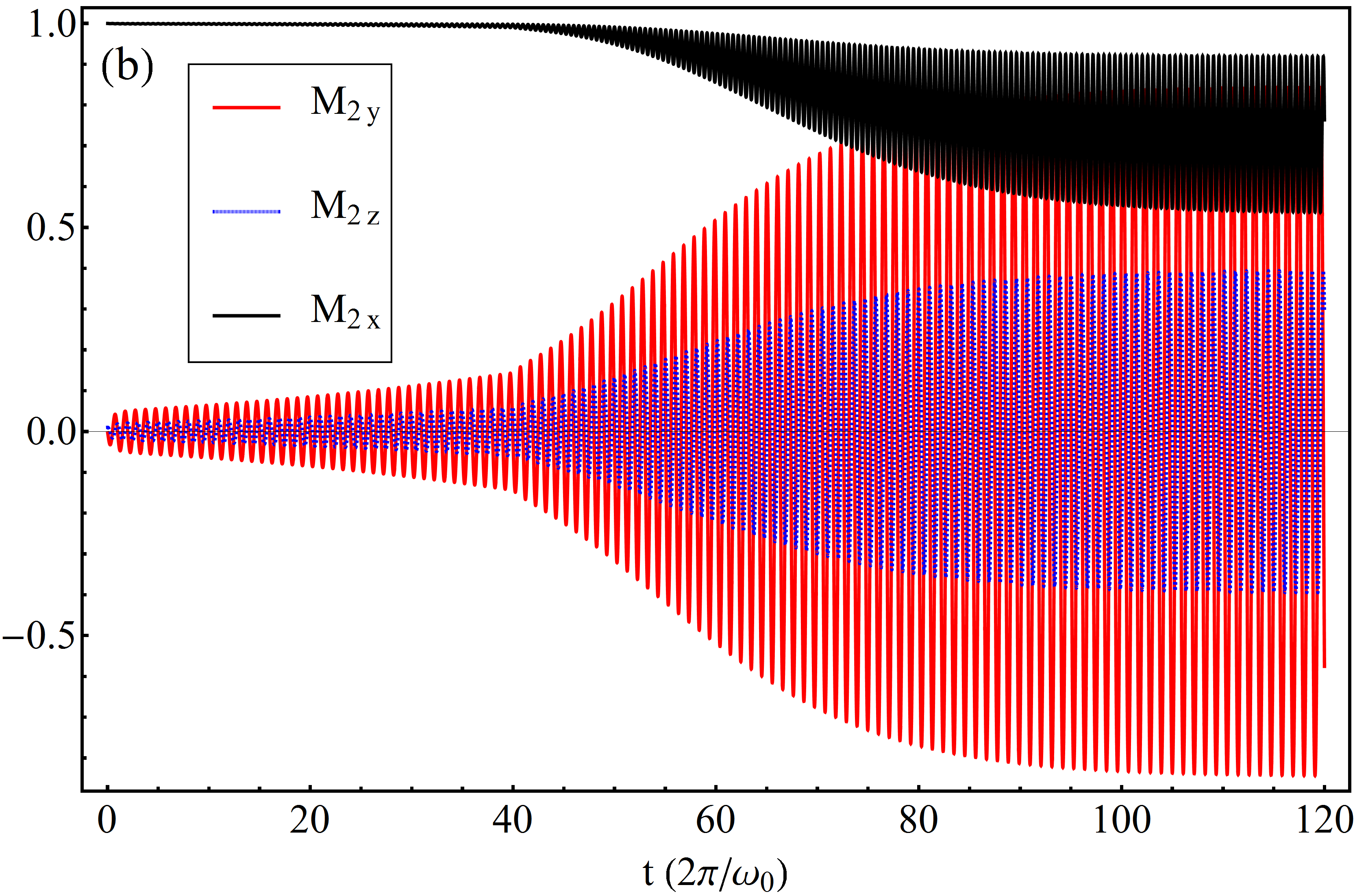}
\caption{(Color online) Magnetization dynamics for the parallel
configuration and currents above the threshold. (a)/(b) the magnetization in
the left/right layer as a function of time in units of $T=2\protect\pi /%
\protect\omega _{0}$. The effective damping is rescaled by letting $%
g_\perp\rightarrow g_\perp 0.005/\protect\alpha $ and $\protect\alpha%
\rightarrow \protect\alpha\, 0.005/\protect\alpha$. The numerical
calculation was carried out by a 4th order Runge-Kutta method with a step
size $\Delta t=T/50$. }
\end{figure}

The non-linear large-angle precession that occurs for currents above the
threshold is not amenable to analytical treatments; however, numerical
calculations can provide some insights. Because the dissipation of YIG is
very low the number of oscillations required to achieve a noticable change
in the precession angle is very large. To speed up the calculations and make
the results more readable we rescale both $g_\perp$ and $\alpha$ by a factor 
$0.005/\alpha$, in this way the effective damping is rescaled. Fig.~\ref%
{fig:omegaparallel} shows the components of the magnetization in the two
layers as a function of time when a large current is switched on for an
initially parallel magnetization along $x$ with a slight canting of $%
M_{i,y}=0.01$ for $i=1,2$. We apply a current $j_{0}^{\mathrm{SH}}/j_{0,%
\mathrm{thr}}^{\mathrm{P,SH}}=110\%$ at $t=0$. For $5T\lesssim t<40T$, the
precession is out of phase, and the amplitude gradually increases. At $t=40T$%
, the applied current is ramped up to $j_{0}^{\mathrm{SH}}/j_{0,\mathrm{thr}%
}^{\mathrm{P,SH}}=130\%$. At $t\sim 60T$, the precession angle is no longer
small, and our previous perturbative treatment breaks down. However, we can
understand that the right layer precesses with a large angle, while the left
layer stays close to the initial equilibrium from the opposite direction of
the interface spin accumulations $\boldsymbol{\mu }_{0}^{\mathrm{SH}}$.

\section{Superlattices}

\label{s:superlattice} A periodic stack of FIs coupled through Ns supports
spin wave excitations propagating in the perpendicular direction. The
coupling between layers is described by Eq.~\eqref{eq:matrix equation};
however, each FI is coupled through the N layers to two neighboring layers.
The primitive unit cell of the superlattice with collinear magnetization is
the FI$|$N bilayer for the parallel configuration (two bilayers in the
antiparallel configuration). For equivalent saturation magnetizations in all
FI layers, we can write for $i\in \mathbb{Z}$ 
\begin{equation}
\mathbf{M}_{i}=s_{i}\hat{\mathbf{x}}+\mathbf{m}_{i},
\end{equation}%
where $s_{i}=1$ for the parallel and $s_{i}=(-1)^{i}$ for the antiparallel
ground state. We can then linearize the expression with respect to the small
parameters $\mathbf{m}_{i}$. An in-plane charge current causes accumulations
of opposite sign in each N layer. The long-wavelength excitations of the
superlattice magnetization can be treated in the continuum limit. Denoting
the total thickness of a unit cell $b=d_{\mathrm{N}}+d_{\mathrm{FI}}$, we
find for the parallel case ($s_{i}=1$) 
\begin{eqnarray}
\partial _{t}\mathbf{m} &=&\hat{\mathbf{x}}\times \left[ \omega _{H}\mathbf{m%
}+\omega _{M}m_{z}\hat{\mathbf{z}}+(\alpha +2\alpha ^{\prime }(1-F^{\prime
}-G^{\prime }))\partial _{t}\mathbf{m}\right.   \notag \\
&&\left. -\alpha ^{\prime }G^{\prime 2}\partial _{t,zz}\mathbf{m}-\omega
_{x}b^{2}\partial _{zz}\mathbf{m}+2j_{0}^{\mathrm{SH}}Gb\partial _{z}\hat{%
\mathbf{x}}\times \mathbf{m}\right] .
\end{eqnarray}%
For $\mathbf{m}=\mathbf{m}_{0}e^{i(\omega t-k_{z}z)}$, the linearized
dispersion relation is 
\begin{eqnarray}
\omega  &=&\sqrt{\omega _{H}(\omega _{H}+\omega _{M})}+\frac{1}{2}(2\omega
_{H}+\omega _{M})\frac{\omega _{x}}{\omega _{0}}b^{2}k_{z}^{2}-2j_{0}^{%
\mathrm{SH}}Gbk_{z}  \notag \\
&&+i\frac{1}{2}(2\omega _{H}+\omega _{M})(\alpha +2\alpha ^{\prime
}(1-F^{\prime }-G^{\prime })+\alpha ^{\prime }G^{\prime }k_{z}^{2}b^{2}).
\end{eqnarray}%
The applied current thus adds a term that is linear in $k_{z}$ to the real
part of the frequency. The direct effect of the SHE now vanishes because the
torques on both sides of any FI cancel. However, when $\mathbf{m}_{0}\neq 
\mathbf{0}$, a net spin current flows normal to the stack, which affects the
dispersion. In the ferromagnetic layers, this phenomenon is equivalent to a
pure strain field on the magnetization and is therefore non-dissipative.
While generating $j_{0}^{\mathrm{SH}}$ causes Ohmic losses, the
magnetization dynamics in this limit do not add to the energy dissipation,
explaining the contribution to $\text{Re}[\omega ].$ In this regime, there
are no external current-induced contributions or instabilities.

Antiferromagnetic superlattices appear to be difficult to realize
experimentally because a staggered external magnetic field would be
required. The unit cell is doubled as is the number of variables in the
equation of motion. Determining the coupling coefficients from Eq.~%
\eqref{eq:matrices} is straightforward but cumbersome and is not presented
here. Naively, one could expect that the SHE-induced torque would act very
differently in the antiferromagnetic case. The SHE acts in a symmetric
manner on the FI($\uparrow$)$|$N$|$FI($\downarrow$) system, stabilizing or
destabilizing both layers simultaneously. However, similarly to the
ferromagnetic superlattice, the direct SHE vanishes also in the
antiferromagnetic superlattice. Each FI is in contact with an N, with spin
accumulations of opposite sign on the left and right side of the interfaces,
which leads to the same cancellation of the direct SHE-induced torque
presented for the ferromagnetic superlattice.

We can also envision a multilayer in which individual metallic layers can be
contacted separately and independently. N$|$FI$|$N structures have been
predicted to display a magnon drag effect through the ferromagnetic film,\cite{Zhang}
i.e. a current in one layer induces an emf in the other one. A
drag effect does also exists in our macrospin model: if we induce dynamics
by a current in one layer by the spin Hall effect, the spin pumping and
inverse spin Hall effect generates a current in the other layer, but only
above a current threshhold. 

With separate contacts to the layers one may drive opposite currents through
neighboring films. In that case, the spin currents absorbed by a
ferromagnetic layer is relatively twice as large as in the FI$|$N bilayer,
thereby reducing the critical currents for the parallel configuration, but
of opposite sign for neighboring magnetic layers. A staggered current
distribution in the superlattice destabilizes the ferromagnetic
configuration, but it can stabilize an antiferromagnetic one even in the
absence of static exchange coupling. This leads to intricate dynamics when
competing with an applied magnetic field.


\section{Conclusions}

\label{s:conclusions} We study current-induced magnetization dynamics in
spin valves and superlattices consisting of insulating magnets separated by
metallic spacers with spin Hall effect. The current-induced torques
experienced by the two magnetic layers in an FI($\uparrow $)$|$N$|$FI($%
\uparrow $) spin valve caused by the spin Hall effect are opposite in sign.
A charge current in N normal to the magnetization this leads to a damping
and an antidamping, stabilizing one and destabilizing the other
magnetization. We calculate the magnetization dynamics when the two layers
are exchange coupled and in the presence of the dynamic exchange coupling
induced by spin pumping. In an antiparallel configuration FI($\uparrow $)$|$N%
$|$FI($\downarrow $) the interlayer couplings play a minor role in the
current-induced effects. The threshold currents at which self-oscillation
occur are higher for parallel than antiparallel spin valves. We predict
interesting current-induced effects for superlattices and multilayers in
which the metallic spacer layers can be individually contacted.

\begin{acknowledgements}
H.S. and A. B. acknowledge support from the Research Council of Norway, project number 216700. This work was supported by KAKENHI (Grants-in-Aid for Scientific Research) Nos. 25247056 and 25220910, FOM (Stichting voor Fundamenteel Onderzoek der Materie), the ICC-IMR, the EU-RTN Spinicur, EU-FET grant InSpin 612759 and DFG Priority Program 1538 "Spin-Caloric Transport" (GO 944/4).\end{acknowledgements}%

\appendix

\section{Matrix Elements}

\label{App: matrix elements} Here, we derive the response coefficients $F$, $%
G$, $F^{\prime }$ and $G^{\prime }$ that determine the torques, depending on
the properties of the normal metal. Let us first discuss the coefficients
related to the torques induced by the SHE. The functions $F$ and $G$ are
extracted from the derivatives of Eq.~\eqref{SH coupling} with respect to
the transverse components of the dynamic magnetizations $\mathbf{m}_{i}$. $F$
governs the SHE-induced torque in one layer due to displaced magnetization
in the same layer and can be computed as 
\begin{equation}
\frac{\partial (\boldsymbol{\tau }_{1}^{\mathrm{SH}})_{y}}{\partial m_{1,y}}=%
\frac{\partial (\boldsymbol{\tau }_{1}^{\mathrm{SH}})_{z}}{\partial m_{1,z}}%
=-\frac{\partial (\boldsymbol{\tau }_{2}^{\mathrm{SH}})_{y}}{\partial
(sm_{2,y})}=-\frac{\partial (\boldsymbol{\tau }_{2}^{\mathrm{SH}})_{z}}{%
\partial (sm_{2,z})}=-Fj_{0}^{\mathrm{SH}}.
\end{equation}%
Thus, 
\begin{eqnarray}
F &=&\frac{\gamma \hbar }{2e^{2}M_{S}d}g_{\perp }\frac{2el_{\mathrm{sf}}}{%
\sigma }\tanh (d_{\text{\textrm{N}}}/2l_{\mathrm{sf}})  \notag \\
&&\left[ 1-\frac{1}{2}\Gamma _{1}\left( d_{\text{N}}/2\right) -\frac{1}{2}%
\Gamma _{2}\left( d_{\text{N}}/2\right) \right] .
\end{eqnarray}%
Similarly, we can identify $G$, which governs the cross-correlation of the
SHE-induced torque in one layer arising from a displaced magnetization in
the other layer from%
\begin{equation}
\frac{\partial (\boldsymbol{\tau }_{1}^{\mathrm{SH}})_{y}}{\partial
(sm_{2,y})}=\frac{\partial (\boldsymbol{\tau }_{1}^{\mathrm{SH}})_{z}}{%
\partial (sm_{2,z})}=-\frac{\partial (\boldsymbol{\tau }_{2}^{\mathrm{SH}%
})_{y}}{\partial m_{1,y}}=-\frac{\partial (\boldsymbol{\tau }_{2}^{\mathrm{SH%
}})_{z}}{\partial m_{1,z}}=Gj_{0}^{\mathrm{SH}},
\end{equation}%
Thus%
\begin{eqnarray}
G &=&\frac{\gamma \hbar }{2e^{2}M_{S}d}g_{\perp }\frac{2el_{\mathrm{sf}}}{%
\sigma }\tanh (d_{\text{\textrm{N}}}/2l_{\mathrm{sf}})  \notag \\
&&\frac{1}{2}\left[ \Gamma _{1}\left( d_{\mathrm{N}}/2\right) -\Gamma
_{2}\left( d_{\mathrm{N}}/2\right) \right] .
\end{eqnarray}%
Torques generated by spin pumping contain terms of the form $\mathbf{\hat{x}}%
\times \mathbf{m}_{i}$ and couple the $y$- and $z$-components of the
magnetization dynamics. We find 
\begin{equation}
\frac{\partial (\boldsymbol{\tau }_{1}^{\mathrm{ST}})_{y}}{\partial \dot{m}%
_{1,z}}=-\frac{\partial (\boldsymbol{\tau }_{1}^{\mathrm{ST}})_{z}}{\partial 
\dot{m}_{1,y}}=\frac{\partial (\boldsymbol{\tau }_{2}^{\mathrm{ST}})_{y}}{%
\partial (s\dot{m}_{2,z})}=-\frac{\partial (\boldsymbol{\tau }_{2}^{\mathrm{%
ST}})_{z}}{\partial (s\dot{m}_{2,y})}=F^{\prime }\alpha ^{\prime },
\end{equation}%
where 
\begin{equation}
2F^{\prime }=\Gamma _{1}\left( d_{\mathrm{N}}/2\right) +\Gamma _{2}\left( d_{%
\mathrm{N}}/2\right) .
\end{equation}%
Similarly, 
\begin{equation}
\frac{\partial (\boldsymbol{\tau }_{1}^{\mathrm{ST}})_{y}}{\partial (s\dot{m}%
_{2,z})}=-\frac{\partial (\boldsymbol{\tau }_{1}^{\mathrm{ST}})_{z}}{%
\partial (s\dot{m}_{2,y})}=\frac{\partial (\boldsymbol{\tau }_{2}^{\mathrm{ST%
}})_{y}}{\partial \dot{m}_{1,z}}=-\frac{\partial (\boldsymbol{\tau }_{2}^{%
\mathrm{ST}})_{z}}{\partial \dot{m}_{1,y}}=G^{\prime }\alpha ^{\prime },
\end{equation}%
where 
\begin{equation}
2G^{\prime }=\Gamma _{1}\left( d_{\mathrm{N}}/2\right) -\Gamma _{2}\left( d_{%
\mathrm{N}}/2\right) .
\end{equation}%
We finally note that some of the coefficients are related:%
\begin{equation}
\frac{G}{G^{\prime }\alpha ^{\prime }}=\frac{1}{\hbar }\frac{2el_{\mathrm{sf}%
}}{\sigma }\tanh (d_{\mathrm{N}}/2l_{\mathrm{sf}}).
\end{equation}


\begin{thebibliography}{99}
\bibitem{Brataas:nmat12} A. Brataas, A. D. Kent, and H. Ohno, Nature Mat., 
\textbf{373} (2012).

\bibitem{Nakayama:prl13} H. Nakayama, M. Althammer, Y.-T. Chen, K. Uchida,
Y. Kajiwara, D. Kikuchi, T. Ohtani, S. Gepr\"{a}gs, M. Opel, S.Takahashi, R.
Gross, G.E. W. Bauer, S. T. B. Goennenwein and E. Saitoh, Phys. Rev.
Lett., \textbf{110} 206601  (2013). 

\bibitem{Kajiwara:nat10} Y. Kajiwara, K. Harii, S. Takahashi, J. Ohe, K.
Uchida, M. Mizuguchi, H. Umezawa, H. Kawai, K. Ando, K. Takanashi, S.
Maekawa, and E. Saito, Nature \textbf{464,} 7269 (2010).

\bibitem{Sandweg:apl10} C. W. Sandweg, Y. Kajiwara, K. Ando, E. Saitoh, and
B. Hillebrands, Appl. Phys. Lett. \textbf{97}, 252504 (2010).

\bibitem{Sandweg:prl11} C. W. Sandweg, Y. Kajiwara, A. V. Chumak, A. A.
Serga, V. I. Vasyuchka, M. B. Jungfleisch, E. Saitoh, and B. Hillebrands,
Phys. Rev. Lett. \textbf{106}, 216601 (2011).

\bibitem{Vilela-Leao:apl11} L. H. Vilela-Leao, C. Salvador, A. Azevedo, S.
M. Rezende, Appl. Phys. Lett. \textbf{99}, 102505 (2011). 

\bibitem{Burrowes:apl12} C. Burrowes, B. Heinrich, B. Kardasz, E. A.
Montoya, E. Girt, Yiyan Sun, Young-Yeal Song, and Mingzhong Wu, Appl. Phys.
Lett. \textbf{100}, 092403 (2012).

\bibitem{Rezende:apl12} S. M. Rezende, R. L. Rodriguez-Suarez, M. M. Soares,
L. H. Vilela-Leao, D. Ley Dominguez, and A. Azevedo, Appl. Phys. Lett. 
\textbf{102}, 012402 (2012).

\bibitem{Xiao:prb10} J. Xiao, G. E. W. Bauer, K.-C. Uchida, E. Saitoh, and
S. Maekawa, Phys. Rev. B \textbf{81}, 214418 (2010).

\bibitem{Yan-Ting:prb13} Y.-T. Chen, S. Takahashi, H. Nakayama, M.
Althammer, S. T. B. Goennenwein, E. Saitoh, and G. E. W. Bauer, Phys. Rev. B 
\textbf{87}, 144411 (2013).

\bibitem{Slonczewski:prb10} J. C. Slonczewski, Phys. Rev. B \textbf{82},
054403 (2010). 

\bibitem{Xingtao:epl11} J. Xingtao, L. Kai, K. Xia, and G. E. W. Bauer, EPL 
\textbf{96}, 17005 (2011).

\bibitem{Kapelrud:prl13} A. Kapelrud and A. Brataas, Phys. Rev. Lett. 
\textbf{111}, 097602 (2013).

\bibitem{Xiao:xxx13} Y. Zhou, H. J. Jiao, Y. T. Chen, G. E. W. Bauer, and J. Xiao,
Phys. Rev. B \textbf{88}, 184403 (2013).

\bibitem{Xiao:prl12} J. Xiao and G. E. W. Bauer, Phys. Rev. Lett. \textbf{108%
}, 217204 (2012).

\bibitem{Heinrich:prl03} B. Heinrich, Y. Tserkovnyak, G. Woltersdorf, A.
Brataas, R. Urban and G. E. W. Bauer, Phys. Rev. Lett. 90, 187601 (2003).

\bibitem{Qiu:rapid92} Z. Q. Qiu, J. Pearson and S. D. Bader, Phys. Rev. B 
\textbf{46}, 8659 (1992).

\bibitem{Beaujour:jap09} J.-M.L. Beaujour, W. Chen, A. D. Kent and J. Z.
Sun, J. Appl. Phys. \textbf{99}, 08N503 (2006).

\bibitem{Tserkovnyak:revmod05} Y. Tserkovnyak, A. Brataas, G. E. W. Bauer
and B. I. Halperin, Rev. Mod. Phys. \textbf{77} 1375 (2005).

\bibitem{Junfleisch:apl13} M. B. Jungfleisch, V. Lauer, R. Neb, A. V. Chumak
and B. Hillebrands, Appl. Phys. Lett. \textbf{103}, 022411 (2013).%

\bibitem{Giancoli:book84} D. Giancoli, "25. Electric Currents and
Resistance". In Jocelyn Phillips. Physics for Scientists and Engineers with
Modern Physics (4th ed.), (2009) [1984]

\bibitem{Serga:jphysd} A. A. Serga, A. V. Chumak and B. Hillebrands, J.
Phys. D: App. Phys. \textbf{43}, 264002 (2010)

\bibitem{Zhang} Steven S.-L. Zhang and S. Zhang, Phys. Rev. Lett. \textbf{109},
096603 (2012).
\end{thebibliography}
\end{document}